\documentclass[superscriptaddress,twocolumn,amsmath,amssymb,showpacs]{revtex4}

\usepackage{graphicx}
\usepackage{dcolumn}
\usepackage{bm}

\begin{document}

\preprint{APS/123-QED}

\title{Anisotropic Release of the Residual Zero-point Entropy in the Spin Ice Compound Dy$_2$Ti$_2$O$_7$: Kagome-ice Behavior}

\author{Ryuji Higashinaka}
\affiliation{
Department of Physics, Kyoto University, Kyoto 606-8502, Japan
}
\author{Hideto Fukazawa}
\affiliation{
Department of Physics, Kyoto University, Kyoto 606-8502, Japan
}
\author{Yoshiteru Maeno}
\affiliation{
Department of Physics, Kyoto University, Kyoto 606-8502, Japan
}
\affiliation{
International Innovation Center, Kyoto University, Kyoto 606-8501, Japan
}

\date{\today}

\begin{abstract}
We report the specific heat and entropy of single crystals of the spin ice compound Dy$_2$Ti$_2$O$_7$ at temperatures down to 0.35 K. We apply magnetic fields along the four characteristic directions: [100], [110], [111] and [112]. Because of Ising anisotropy, we observe anisotropic release of the residual zero-point entropy, attributable to the difference in frustration dimensionality. In the high magnetic field along these four directions, the residual entropy is almost fully released and the activation entropy reaches $R$ln2. However, in the intermediate field region, the entropy in fields along the [111] direction is different from those for the other three field directions. For the [111] direction the frustration structure changes from that of three-dimensional(3D) pyrochlore to that of two-dimensional(2D) Kagome-like lattice with constraint due to the ice rule, leading to different values of zero-point entropy.
\end{abstract}

\pacs{65.40.Ba, 65.40.Gr, 75.25.+z, 75.50.Lk}
\maketitle
\begin{figure}[btp]
\includegraphics[width=0.9\linewidth]{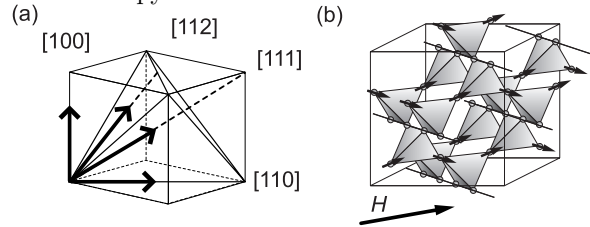}
\caption{\label{fig:direction4} (a) Four characteristic directions in which the magnetic field is applied. (b) 1D chain of spins perpendicular to the field on a pyrochlore lattice in the [110] field direction.}
\end{figure}
\begin{figure}[btp]
\includegraphics[width=0.8\linewidth]{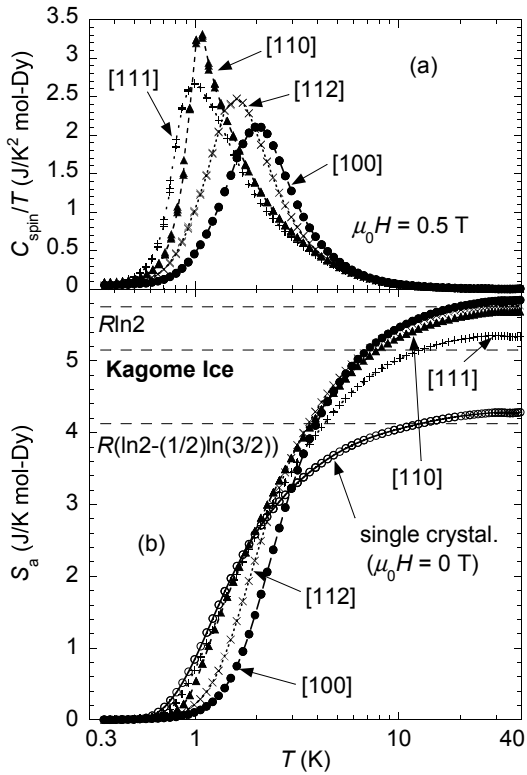}
\caption{\label{fig:h0.5T} The spin contributions to (a) specific heat and (b) activation entropy of single crystalline Dy$_2$Ti$_2$O$_7$ at 0.5 T along four characteristic directions: [100], [110], [111] and [112], and at 0 T(entropy only). The dotted line for the Kagome ice is based on Pauling's approximation.}
\end{figure}

 Recently, there has been increasing attention in the physics of geometrical frustration in pyrochlore oxides $A_2 B_2$O$_7$. The spin ice behavior \cite{Review} has been observed in Ho$_2$Ti$_2$O$_7$ \cite{Harris97,LGHarris,Bramwell01,NeutronHoTi,Petrenko02}, Dy$_2$Ti$_2$O$_7$ \cite{RamirezN,Fukazawa02,Fennell01,Matsuhira01,Snyder01} and Ho$_2$Sn$_2$O$_7$ \cite{Kadowaki02}. Since there is no long-range magnetic ordering of the rare-earth moments in these materials at least down to 50 mK \cite{Harris97,Fukazawa02}, the ground state is believed to be macroscopically degenerate. In fact, the observed residual entropy \cite{RamirezN,Higashinaka02} shows an excellent agreement with the expected zero-point entropy of (1/2)$R$ln(3/2) \cite{Pauling45}. \\
\quad In these materials, the $A$-site ions constitute a three-dimensional (3D) network of corner-shared tetrahedra (the pyrochlore lattice). Because of the crystal field effect, the magnetic moments of the $A$-site ions, such as Dy$^{3+}$ and Ho$^{3+}$, have Ising anisotropy along the local $\langle$111$\rangle$ direction, which points to the center of the tetrahedron from a vertice. Owing to this Ising anisotropy, the spin responses to magnetic fields are very anisotropic. In measurements on polycrystalline samples, there are specific-heat peaks at field-independent temperatures of 0.34 K, 0.47 K and 1.12 K; it was speculated that these peaks may be due to ordering of spins with their Ising axes perpendicular to the field \cite{RamirezN}. Moreover, owing to the difference in the spin configurations in fields along different directions, the process of releasing the residual zero-point entropy should be qualitatively different reflecting the structure of frustration. However, up to present there are few studies of the entropy release using single crystals \cite{Higashinaka02,Matsuhira02,HiroiC}. \\
\quad In this paper, we report the specific heat and entropy of single crystalline Dy$_2$Ti$_2$O$_7$ in magnetic fields along four characteristic directions, [100], [110], [111] and [112]. To the best of our knowledge, relevant data for the [110] and [112] directions have not been published previously. We indeed observed an anisotropic process of releasing residual entropy for fields along different directions. We will discuss this anisotropic behavior in terms of a qualitative difference in the dimensionality of frustration structures. \\
\quad The single crystals of Dy$_2$Ti$_2$O$_7$ and Eu$_2$Ti$_2$O$_7$ used in this work were grown by a floating zone method \cite{Fukazawa02}. We measured the specific heat between 0.35 and 40 K and in fields up to 5 T by a relaxation method using a commercial calorimeter (Quantum Design, model PPMS). To obtain reliable entropy values, it was essential to measure the specific heat of the sample holder (addenda) at each corresponding magnetic field. In this way, our data presented here does not involve data manipulation to compensate for the field dependence of the background. \\
\quad The sample size was approximately $2.0 \times 2.0 \times 0.06 $mm$^{3}$ and the mass was about 1.7 mg. In order to avoid field inhomogeneity due to the demagnetization effect, we cut the crystal along the (1\=10) plane. All the four characteristic directions of the cubic lattice, [100] (equivalent to [001]), [110], [111] and [112] (Fig.~\ref{fig:direction4} (a)), lie in the surface plane of the plate-like sample. This configuration allows us to examine the anisotropy based on the data taken from the same crystal. \\
\quad Because Dy$_2$Ti$_2$O$_7$ is an insulator, the total specific heat, $C_{{\rm total}}$, can be expressed as
\begin{equation}
C_{{\rm total}} = C_{{\rm spin}} + C_{{\rm phonon}},
\end{equation}
where $C_{{\rm spin}}$ and $C_{{\rm phonon}}$ represent electronic-spin and phonon contributions, respectively. The nuclear Schottky term is negligible in this compound. In order to estimate the phonon contribution, we also measured the specific heat of Eu$_2$Ti$_2$O$_7$. Because of the similarities of the lattice parameters and the atomic mass of the $A$-site ions, and of the nonmagnetic ground state and insulating property of Eu$_2$Ti$_2$O$_7$, we approximate $C_{{\rm phonon}}$ of Dy$_2$Ti$_2$O$_7$ with the total specific heat of Eu$_2$Ti$_2$O$_7$ with a small correction for the difference in $A$-site mass, ($M_{{\rm Dy}}/M_{{\rm Eu}}$) $C'_{{\rm total}}$(Eu$_2$Ti$_2$O$_7$). Thus the spin contribution is approximated as $C_{{\rm spin}}({\rm Dy}_2{\rm Ti}_2{\rm O}_7) \cong C_{{\rm total}}({\rm Dy}_2{\rm Ti}_2{\rm O}_7) - (M_{{\rm Dy}}/M_{{\rm Eu}}) C'_{{\rm total}}({\rm Eu}_2{\rm Ti}_2{\rm O}_7)$. Here $M_{{\rm Dy}}$ and $M_{{\rm Eu}}$ represent molar mass of Dy$_2$Ti$_2$O$_7$ and Eu$_2$Ti$_2$O$_7$, respectively. We estimate the activation entropy $S_{a}(T)$ by integrating $C_{{\rm spin}}/T$ from 0.35 K to $T$ of up to 40 K:
\begin{equation}
S_{a}(T) = \int^{T}_{0.35{\rm \ K}} \frac{C_{{\rm spin}}}{T} {\rm d}T
\end{equation}
We deduce the residual entropy, $S_0$, as $R{\rm ln}2 - S_{a}(T_{1})$, using the activation entropy at high temperature $T_{1} \gg J_{{\rm eff}}$ where $J_{{\rm eff}}$ is the effective interaction between the nearest-neighbor spins. In Dy$_2$Ti$_2$O$_7$, the main contribution to $J_{{\rm eff}}$ = 1.11 K is the dipolar interaction and is ferromagnetic (FM) \cite{Fukazawa02,Dipolar,Melko01}. \\
\quad In Fig.~\ref{fig:h0.5T} (a) we show the spin component of the specific heat measured after field-cooling at 0.5 T along the four characteristic directions. It is estimated that the first excited level under the crystal field is nearly 150 K \cite{JanaDy}, so that the total spin entropy involved at low temperature is $R$ln2. The Zeeman energy of the applied field for flipping a spin is expressed as $\Delta E_{{\rm Z}}$ = 2$g_J J \mu_{{\rm B}}(\mu_0 H) \times {\rm cos} \theta$, where $g_J$ is the Land${\rm \grave{e}}$ g factor, $J$ is the total angular moment of Dy$^{3+}$ ion \cite{JanaDy}, $\mu _{{\rm B}}$ is the Bohr magneton, and $\theta$ is the relative angle between the local $\langle$111$\rangle$ and the field direction. The Zeeman energy for a field of $\mu _0 H=$ 0.5 T along the Ising axis is 6.8 K. In a pyrochlore lattice, the four spins in each tetrahedron have Ising axes along local $\langle$111$\rangle$ directions and thus point to different directions; hence the average value of the Zeeman energy is smaller than this value. In fact at 0.5 T, the Zeeman energy for one of the four spins exceeds the internal field governing the ice rule, $4J_{{\rm eff}}$. It should be noted that Fig.~\ref{fig:h0.5T} (b) clearly shows that only for the [111] field direction there still remains residual entropy of $S_0 \approx$ 0.44 J/K mol-Dy. \\
\quad Let us first examine the origins of the specific heat peaks based on the spin ice model \cite{Higashinaka02}. In fields along the [100] direction, the stable spin configuration is uniquely determined as two-spins in and two-spins out (2-in 2-out), satisfying the ice rule. Moreover, all the spin components parallel to the field are equivalent. Therefore there should be only one peak in $C_{{\rm spin}}(T)/T$ associated with the thermal excitation across the energy gap determined by $|4J_{{\rm eff}}$ + $\Delta E_{{\rm Z}}|$. Since cos$\theta$ is $1/\sqrt{3}$ the additional Zeeman energy term at 0.5 T is 3.9 K. This quantitatively explains the observed linear increase of the peak temperature with the field strength in a high field region \cite{Higashinaka02}. In fields lower than 2 T, a changeover to behavior with a nearly field-independent specific heat peak attributable to $J_{{\rm eff}}$ is observed. \\
\quad In the [110] field direction, two of the four spins in each tetrahedron is determined as 1-in 1-out with cos$\theta$ = $\sqrt{2}/\sqrt{3}$, giving the expected additional Zeeman energy term of 5.4 K. The other two spins are perpendicular to the applied field; their directions are not determined by the external field. Thus the former spins should give rise to a Schottky peak with a peak temperature increasing with the field, whereas the latter spins may lead to specific heat anomaly associated with ordering of these sublattice spins due to their interactions. Since the peak position shown in Fig.~\ref{fig:h0.5T} (a) is unchanged with the field strength for $H \|$ [110], we attribute the field-independent peak at 1.1(1) K observed in polycrystalline samples \cite{RamirezN,Higashinaka02} to the ordering of these two free spins. \\
\quad In the [111] field direction, one of the four spins is parallel to the field and most effectively stabilized by the external field. However for the other three spins, the nearest neighbor interaction dictating the ice rule competes with the Zeeman energy of the external field, because the latter demands the 1-in 3-out configuration. Therefore the energy to flip one of these three spins is the {\it difference} between those two interactions. The energy gap becomes $|4J_{{\rm eff}}$ - $\Delta E_{{\rm Z}}|$ with cos$\theta$ = $1/3$, which reaches zero at 1.0 T. In lower fields, the stable spin configuration is the ice-rule 2-in 2-out state because it is dominated by the spin-spin interaction. However, in higher fields the ice rule breaks down and the stable configuration changes to 1-in 3-out. This crossover process has been quantitatively confirmed in the magnetization measurements \cite{Fukazawa02}. \\
\quad In another characteristic field direction, the [112], the configuration of three of the four spins are specified as 1-in 2-out. The value of cos$\theta$ is $2 \sqrt{2}/3$ for one and $\sqrt{2}/3$ for the other two, giving the expected additional Zeeman energy at 0.5 T of 6.4 K and 3.2 K, respectively. The other spin in a tetrahedron is perpendicular to the field. Because of the internal interaction, however, this spin points outward to satisfy the ice rule. Therefore the energy difference between up and down states for this spin depends not on the external field strength, but only on the internal field. There is a possibility that this field-perpendicular spin is the origin of some of the other field-independent peaks observed in polycrystalline specific heat. However, for the [112] field direction we observed only one peak, which depends on $H$ (data not shown). \\
\quad Let us now compare in Fig.~\ref{fig:h0.5T} (b) the spin contribution to the entropy $S_{a}$ at 0.5 T along the four characteristic directions and that of a single crystal at 0 T. Let us first examine the observed entropy release at low temperatures. In the low temperature region, the entropy is most actively released for the [111] field direction, because the energy gap is reduced by the external field as explained above. In the other characteristic field directions, external fields do not compete with the ice rule; they stabilize some particular spin configurations within the spin-ice manifold. It is thus the number of field-perpendicular spins which mainly determine the activation entropy at low temperatures, since these spins are not further stabilized by the magnetic field. The numbers of such spins are two, one, and zero for $H$ // [110], [112], and [100], respectively. The activation entropy at low temperatures indeed decreases in this order as shown in Fig.~\ref{fig:h0.5T} (b). \\
\quad Next, we focus on the values of the activation entropy in the high temperature region. The difference from $R$ln2 represents the residual entropy, which is not released by the external field, and are intimately related to the frustration structure. In the [111] field direction, the nearest neighbor interaction governing the ice rule is stronger than the external field for three of the four spins up to the field of 1.0 T. Thus at 0.5 T the frustration still remains. These three frustrated spins constitute a Kagome lattice (Fig.3). Viewed from the [111] direction, the pyrochlore lattice consists of a layered stacking of a triangular lattice and a Kagome lattice (a pyrochlore slab). The frustration on a 3D pyrochlore changes to that on a 2D Kagome lattice. The ground state of an ordinary Kagome lattice has greater residual entropy than that of pyrochlore lattice \cite{Kano53,Anderson56}. However, in the present case the residual entropy of this Kagome lattice is smaller than that of a pyrochlore lattice because spins on this Kagome lattice have the Kagome-ice rule constraint that the "up triangle" satisfies 2-in 1-out and the "down triangle" satisfies 1-in 2-out (Fig.3). Thus, this state may be called as "Kagome ice" \cite{Kagomeice1}. Its residual entropy is estimated as $\frac{3}{4}R{\rm ln}3^{\frac{N}{3}} ( \frac{4}{9} )^{\frac{N}{3}} \fallingdotseq 0.598$ J/K mol-Dy by Pauling's method \cite{Entropy}. Udagawa $et$ $al.$ calculated the exact value for the Kagome ice model to be 0.672 J/K mol-Dy \cite{Udagawa02}. \\
\quad As shown in Fig.2(b), in fields along the other three direction the residual entropy is fully released ($S_{{\rm a}}$ = 5.78 $\pm$ 0.08 J/K mol-Dy) but along [111] direction, $S_{{\rm a}}$ = 5.35 J/K mol-Dy. Thus in this study the residual entropy is evaluated as $S_{0}$ = 0.44 $\pm$ 0.12 J/K mol-Dy. This value is somewhat smaller than the theoretical expectations. At 1 T along the [111] direction, the effective Zeeman energy overcomes the energy scale of the nearest neighbor FM interaction and spin configuration becomes the ice-rule breaking 1-in 3-out. Therefore the residual entropy should vanish and this expectation agrees with our experimental results (Fig.4). Note that in refs. \cite{Matsuhira02,HiroiC}, they reported for the Kagome-ice residual entropy of 0.8 J/K mol-Dy and 0.65 J/K mol-Dy after subtracting the extrinsic background estimated by a quadratic fitting. \\
\quad In zero field, the role of long-range dipolar interaction is known to be quite important for quantitative description of the spin ice \cite{Dipolar}. For the Kagome ice realized in the [111] field direction, its lower dimensionality should make less important the long-range dipolar interaction within the Kagome lattice, since the increase in the number of lattice sites with the distance is not as rapid as to compensate for the decay in the dipolar interaction. We note, however, that the dipolar field from the ordered moment of the triangular lattice provides an additional molecular field not included in the "Kagome-ice rule". Such dipolar field at a Lattice-lattice site amounts to a correction of -19$\%$ to the nearest-neighbor contribution if the contribution from up to the fourth-nearest-neighbor sites on the triangular lattice is included. \\
\quad In the characteristic field directions other than [111], the zero-point entropy is more readily released. In the [100] and [112] field directions, the spins are uniquely determined in the 2-in 2-out configuration without degeneracy. In the [110] field direction, the structure of frustration changes to a one-dimensional (1D) chain without macroscopic degeneracy as shown in Fig.1 (b). In reality, to overcome the spin freezing setting in below $\sim 1$ K external field of $\sim 0.1$ T is needed to remove the spin-ice residual entropy even in the field-cooling process. Fig. 2 (b) indicates that 0.5 T is indeed sufficient to remove the residual entropy. \\
\quad In conclusion, we presented anisotropic release of the zero-point entropy of single crystalline Dy$_2$Ti$_2$O$_7$, ascribable to the qualitative difference in the structure of frustration. In the [111] field direction, we consider this system as a dipolar Kagome ice, in which the spins frustrate on a 2D Kagome lattice with ice-rule constraint \cite{Kagomeice2} in the presence of additional dipolar field from the ordered moments on the triangular lattice. We estimate the residual entropy by considering the frustration only on the Kagome lattice, but the role of the dipolar field from the triangular-lattice slab is expected to be important in the value of the residual entropy, as well as in the process of entropy release. Such difference between the Kagome-ice behavior and the "dipolar Kagome-ice" behavior is worth further attention in the future studies.\\
\quad We observed a nearly field-independent peak in the specific heat in fields along the [110] direction. This is most probably the origin of one of the field-independent peaks observed in polycrystalline samples at 1.1 K \cite{RamirezN,Higashinaka02}. In the [112] field direction, one of the four spins is perpendicular to the external field. Although there is a possibility that ordering of this "free" spin is the origin of another field-independent peak observed in polycrystals, we did not observe any sign of such a peak in the present measurements down to 0.35 K. \\
\quad We acknowledge helpful discussion with M.J.P. Gingras, R.G. Melko, H. Yaguchi, S. Fujimoto D. Yanagishima and M. Udagawa. We would also like to thank T. Ishiguro for his support in many aspects. H.F. is grateful for the financial support from JSPS. This work has been supported by Grants-in-Aid for Scientific Research from the Japan Society for Promotion of Science and from the Ministry of Education, Culture, Sports, Science and Technology.

\begin{figure}[btp]
\includegraphics[width=0.65\linewidth]{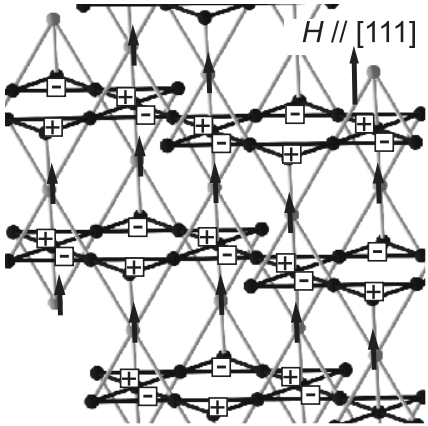}
\caption{\label{fig:kagome} The Kagome lattice in a pyrochlore lattice. Plus and minus signs indicate that along the field direction the net moment of the three spins constituting a triangle is positive and negative, respectively, by the ice rule.}
\end{figure}

\begin{figure}[btp]
\includegraphics[width=0.75\linewidth]{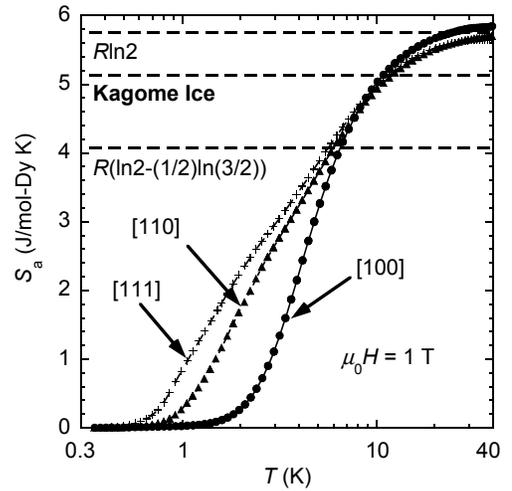}
\caption{\label{fig:hpara} The spin contribution to the activation entropy of single-crystalline Dy$_2$Ti$_2$O$_7$ in field along three characteristic directions at 1 T. The dotted line for the Kagome ice is based on Pauling's approximation.}
\end{figure}


\begin{thebibliography}{25}
\bibitem[{Kag({\natexlab{a}})}]{Kagomeice1}
\bibinfo{note}{The term ''Kagome ice'' is due to Z. Hiroi; private
  communication}.
\bibitem[{Ent()}]{Entropy}
\bibinfo{note}{When there are $N'$ spins on the Kagome lattice, the number of
  the up triangle is $\frac{N'}{3}$. Because there are three configurations on
  each up triangle, the number of state is $3^{\frac{N'}{3}}$. Next, to satisfy
  the down triangle's constraint, we multiply the correction term
  $(\frac{4}{9})^{\frac{N'}{3}}$. Finally, $N'$ is related to the total number
  of spins $N$ by $N' = \frac{3}{4}N$.}
\end{thebibliography}
\end{document}